\documentclass[twocolumn,showpacs,amsmath,amssymb,prl,nofootinbib,superscriptaddress]{revtex4}

\usepackage{amsmath,amsfonts,amssymb,bm,graphicx,xcolor}
\usepackage[color,matrix,arrow,all,2cell]{xy}
\usepackage[colorlinks=true,linkcolor=blue,urlcolor=blue,citecolor=blue,pdfusetitle]{hyperref}

\newcounter{sm}
\newcommand{\sml}[2]{\refstepcounter{sm}\label{#1}\vspace{.2cm} \emph{\arabic{sm}. #2}\vspace{.2cm}}

\newcommand{\e}{\overline{\ell}}
\newcommand{\ba}{\begin{array}}
\newcommand{\ea}{\end{array}}
\newcommand{\bs}[1]{\boldsymbol{#1}}
\newcommand{\up}{\uparrow}
\newcommand{\down}{\downarrow}
\newcommand{\cur}{c}
\newcommand{\R}{S}

\begin{document}

\title{Beat of a current}

\author{Pedro E. Harunari}
\email{pedro.harunari@uni.lu}
\affiliation{Instituto de F\'isica da Universidade de S\~ao Paulo, 05314-970 S\~ao Paulo, Brazil} 
\affiliation{Department of Physics and Materials Science, University of Luxembourg, Campus
Limpertsberg, 162a avenue de la Fa\"iencerie, L-1511 Luxembourg (G. D. Luxembourg)}

\author{Alberto Garilli}
\affiliation{Department of Physics and Materials Science, University of Luxembourg, Campus
Limpertsberg, 162a avenue de la Fa\"iencerie, L-1511 Luxembourg (G. D. Luxembourg)} 

\author{Matteo Polettini}

\affiliation{Department of Physics and Materials Science, University of Luxembourg, Campus
Limpertsberg, 162a avenue de la Fa\"iencerie, L-1511 Luxembourg (G. D. Luxembourg)} 

\begin{abstract}
The fluctuation relation, a milestone of modern thermodynamics, is only established when a set of fundamental currents can be measured. Here we prove that it also holds for systems with hidden transitions if observations are carried ``at their own beat,'' that is, by stopping the experiment after a fixed number of visible transitions, rather than the elapse of an external clock time. This suggests that thermodynamic symmetries are more resistant to the loss of information when described in the space of transitions.
\end{abstract} 



\maketitle

\begin{figure}[t]
\vspace{-.4cm}
\includegraphics[width=\columnwidth]{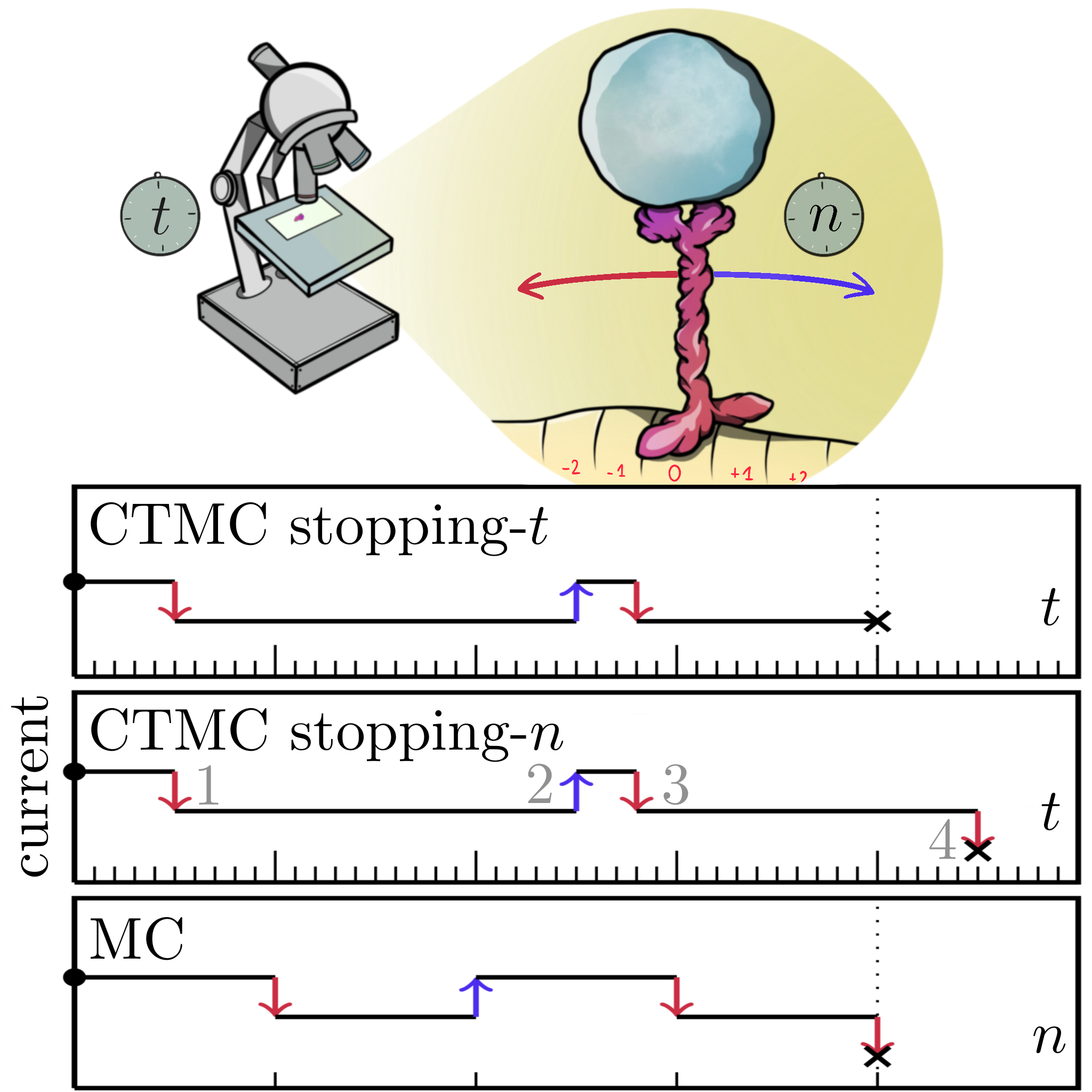}
\caption{When monitoring a current, trajectories are usually collected at the beat of an external clock (continuous-time Markov chain stopping-$t$). A new paradigm, allowing for the fluctuation relation recovery, involves ending the observation at the internal notion of time given by number of transitions \(n\) (continuous-time Markov chain stopping-$n$). Furthermore, the notion of time can be completely washed-away by a Markov chain in transition space (Markov chain).}
\label{fig:fig1}
\end{figure}

Consider the ``symbolism of atomic measurements'', as Schwinger called quantum mechanics \cite{schwinger2003quantum}:  transitions in the energy spectrum of atoms were then only visible through spectral lines, i.e. the emission of photons. Or else, consider a chemical reactor fed by the in- and out-take of some controlled species: while flows can be monitored, the abundance of the reactants is only accessible by scanning with devices that involve internal degrees of freedom -- e.g. magnetic, vibrational, electronic (NMR, UV/Vis and infrared \cite{wilson2016autonomous,amano2021catalysis,sorrenti2017non,semenov2016autocatalytic} spectroscopy). Yet again, as in Fig.\;\ref{fig:fig1}, consider myosins carrying cargoes on actin filaments: their motion can be monitored via imaging techniques, but not their ATP-ADP metabolic cycle \cite{mallik2004cytoplasmic, vale1985identification, nishiyama2002chemomechanical}.

The physics of open systems is a discourse about transitions and transformations. However, our modern understanding based on continuous-time Markov chains is tightly bound to notions of the system's internal state. Take the fluctuation relation, the most encompassing result about nonequilibrium systems, stating that for currents $\bs{\cur}$ cumulated up to some stopping time $\tau$ the log-ratio of their positive to negative probabilities is linear
\begin{align}
\log \frac{p_\tau(\bs{\cur})}{p_\tau(-\bs{\cur})} = \bs{f} \cdot \bs{\cur}.  \label{eq:fluctuation relation}
\end{align}
The above relation holds at times $\tau = t$ beat by an external clock (upon a proper choice of preferred initial distribution \cite{polettini2014transient}, or asymptotically) only if the observer has access to all currents and forces in the system's state space, up to boundary contributions. Instead, it does not generally hold if some of the currents are not visible.

Our main result is that a local current $\cur = n_\up - n_\down$, that is the number of times a certain transition denoted $\up$ occurs minus that of the opposite transition $\down$, obeys the fluctuation relation when counted ``at its own beat'', namely the number of times $\tau \equiv n = n_\up + n_\down $ that either $\up$ or $\down$ are performed, regardless of what happens within the system in the meanwhile.

The second main contribution is the introduction of the formalism of Markov chains in the space of transitions, rather than states, which we prove to correctly describe the statistics of observables at total number of visible transitions. We illustrate this in Fig.~\ref{fig:fig1}. Whilst the letter is self-contained, we refer with superscripts$^i$ to Supplemental Material.

\paragraph{Setup.} We work with autonomous continuous-time Markov chains $x(t)$ from $t=0$ to some stopping time $\tau \in [0,+\infty)$ with rates $r(x|y)$ of jumping from $y$ to $x$. All probabilities here and in the following can in principle be derived from a well-known path probability density $p(\{x(t), t \in [0,\tau] \})$ that in simulations will be produced by the Doob-Gillespie algorithm \cite{weber2017master}. The state space can be depicted as a graph with states as nodes and transitions as directed links, for example:

\begin{equation}
\label{eq:model}
\includegraphics[trim={0 0.3cm 0 0.6cm}, width=0.28\columnwidth]{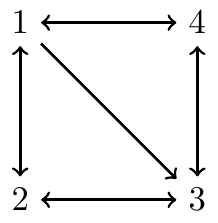}
\end{equation}

Notice that not all transitions need to be reversible. Removal of links $1 \leftrightarrow 2$ and $3 \leftrightarrow 4$ makes the graph into a tree, which supports no stationary current \cite{zia2007probability}. We will instead focus on one link only, belonging to some cycle. Without loss of  generality we take $\up \,= 1 \gets 2$ and $\down \, = 2 \gets 1$ as our visible transitions, on the assumption that there are no other mechanisms connecting $1$ and $2$ directly. We further assume {\it hidden irreducibility}, i.e. the existence of a non-zero probability path between any pair of states not containing visible transitions.

For a physical picture, these transitions could be associated with measurable emission and absorption of photons of energy $\varepsilon$ with a thermal bath at inverse temperature $\beta$. Local detailed balance then grants that
\begin{align}
\frac{r(1|2)}{r(2|1)} = \exp \beta \varepsilon.
\end{align}
We assume non-degeneracy, that is $1\leftrightarrow 2$ is the only transition exchanging photons of that energy, and that the temperature can be regulated. 

We define $R$ as the rate matrix with entries $R_{x,y} := r(x|y) - \delta_{x,y} r(y)$, with $r(y) = \sum_{x} r(x|y)$ the exit rate out of state $y$, and $\delta$ Kronecker's. Consider the induced chain  of states visited by the process $\bs{x} = \{x_m \gets x_{m-1} \gets \ldots \gets x_0\}$, and let the taboo function $\theta(\bs{x})$ be zero whenever any two consecutive states are either $(1,2)$ or $(2,1)$, otherwise it is $1$. We define the taboo matrix $T$ with entries $T_{x,y} := \theta(x,y) = 1 - \delta_{x,1} \delta_{y,2} - \delta_{y,1} \delta_{x,2}$ and the survival rate matrix $\R := R \circ T$, where $\circ$ is component-wise Hadamard multiplication. In other words, the survival  rate matrix is identical to $R$ but for $\R_{1,2} = 0 = \R_{2,1}$.

\paragraph{Trans-transition probabilities.}

Letting $p_t(\bs{n}|x)$ be the probability that, starting from $x$, one observes total numbers $\bs{n} = (n_{\up}, n_{\down})$ of transitions $\up, \down$ up to time $t$, the survival probability of not performing any visible transition is found to be${}^1$ \cite{sekimoto2021derivation,harunari2022learn}
\begin{align}
p_t(\bs{0}|x) = \sum_y \left[\exp t \R \right]_{y,x}.
\end{align}
Taking minus the time derivative we find
\begin{align}
- \frac{d}{dt}  p_t(\bs{0}|x)  =  r(1|2) [ e^{t \R} ]_{2,x} + r(2|1) [ e^{t\R} ]_{1,x}, \label{norm}
\end{align}
where we used the fact that columns of $R$ add up to zero. On the right-hand side the two contributions can be proven (cf. Appendix A of \cite{harunari2022learn}) to be respectively the rates at which $\up$ or $\down$ are performed for the first time in the time interval  $[t, t +dt)$, defining a renewal Markov process in the space of transitions (or,  if the transitions are thought to be prolonged for the whole duration of the interval, a semi-Markov process as in Ref.\,\cite{martinez2019inferring}). Integrating the first contribution over time we find
\begin{align}
p(\up|\up) := - r(1|2) \, [\R^{-1}]_{2,1} \label{eq:upup}
\end{align}
where we used the fact that the eigenvalue of $\R$ with largest real-part is negative${}^2$ \cite{suzumura1983perron}, implying $\lim_{t \to \infty} e^{t\R} = 0$. Equation \,(\ref{eq:upup}) is indeed the probability that the next transition is $\up$, given that the previous was $\up$, by 1) the strong Markov property that grants that the process in state space stays Markov for any notion of stopping time (in this case, that of the next transition), and 2) by the fact that no two microscopic transitions contribute to the same observable. We dub this and other similar expressions${}^3$ $p(\ell | \ell')$, where $\ell$ (for ``link'') denotes a generic transition $\in \{\up, \down\}$, the {\it trans-transition} probabilities. A useful formula for their interpretation is${}^4$
\begin{align}
- [\R^{-1}]_{x,x_0} = \frac{1}{r(x)} \sum_{\bs{x} \;:\; x_0 \rightsquigarrow x} \theta(\bs{x}) \, p(\bs{x} | x_0),
\end{align}
where the sum runs over all trajectories, of any length, that go from $x_0$ to $x$. $p(\bs{x} | x_0)$ is the probability of the induced Markov chain that can be obtained from transition rates, and \( \theta (\bs{x}) \) filters the trajectories that include visible transitions. Notice that, by hidden irreducibility, trans-transition probabilities are positive.

\paragraph{Markov chain in transition space.}

We can arrange trans-transition probabilities in a trans-transition matrix
\begin{align}
P := \left(\ba{cc} p(\up|\up) & p(\up|\down) \\ p(\down|\up) & p(\down|\down) \ea \right). \label{eq:Pkappa} 
\end{align}
By normalization of Eq.\,(\ref{norm}) with respect to $t$, columns of $P$ add up to unity. Therefore $P$ is a discrete-time transition matrix in the following space of transitions:
\begin{equation}
\includegraphics[width=.4\columnwidth]{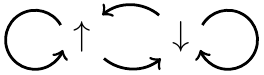}
\end{equation}

Thus, the sequence of visible transitions is a Markov chain in transition space, which by hidden irreducibility is fully connected. Notice that here the Markov property is preserved by lifting the observable process into a different space. Other decimation procedures anchored on states typically break Markovianity, which is only recovered in the limit of time-scale separation \cite{bo2017multiple,esposito2012stochastic}.

Now consider the probability $p_n(\ell)$ that the $n$-th transition is $\ell$. Collect them in a vector $\vec{p}_n$ and let $\mathtt{s}(\ell)$ and $\mathtt{t}(\ell)$ denote the source and target states of the transition. Then, given the initial state probability $q_0(x)$ of being in $x$ at clock time $t = 0$, once obtained the probability of the first transition as
\begin{align}
p_1(\ell) = - r(\mathtt{t}(\ell)|\mathtt{s}(\ell))  \sum_{x} [ \R^{-1} ]_{\mathtt{s}(\ell),x} q_0(x), \label{eq:init}
\end{align}
which is also normalized${}^5$, we can further evolve the process in transition space by $\vec{p}_n = P^{n-1} \vec{p}_1$. Notice that the Markov chain's ``beat'' is that of the occurrences of visible transitions, rather than the clock time $t$ or the total number of jumps in state space usually considered.

\paragraph{Paths and time reversal.}

We can depict induced chains $\bs{x}$ as walks in the above graph in Eq.\,(\ref{eq:model}), e.g.

\begin{align}
\begin{aligned}
\includegraphics[trim={0 0.2cm 0 0.4cm},width=.6\columnwidth]{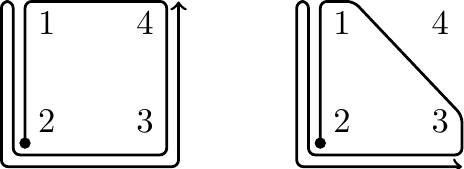}
\end{aligned}
\label{eq:pathsstates}
\end{align}

Searching for the visible transitions in $\bs{x}$ we can map paths in state space into paths in transition space $\bs{x} \mapsto \bs{\ell} = \{\ell_{n} \gets \ell_{n-1} \gets \ldots \gets \ell_{1} \}$. Notice that the two example paths above correspond to the same path $\up\up\down$ in transition space, on the left-hand side:

\begin{align}
\begin{aligned}
\includegraphics[trim={0 0.2cm 0 0.6cm}, width=.7\columnwidth]{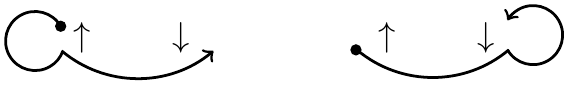}
\end{aligned}
    \label{eq:picpic}
\end{align}
On the right-hand side is the time-reversed path in transition space,  which upon the hidden irreducibility hypothesis always exists, despite the fact that not all state-space paths afford a time-reversed process, e.g. the second in Eq.\,(\ref{eq:pathsstates}). Notice that we do not only invert the order of the transitions, but also flip their direction. Time reversal is involutive, and therefore bijective. 

The probability of transition path $\bs{\ell}$ is
\begin{align}
p(\bs{\ell}) = p_1(\ell_1) \prod_{k = 1}^{n-1} p(\ell_{k+1} | \ell_{k}).
\end{align}
We now compare it to that of its time-reversed, both sampled from the same initial distribution, by taking their ratio. The time-reversed of $p(\ell | \e)$ is itself, therefore all such terms cancel out and we are left with
\begin{align}
\frac{p(\bs{\ell})}{p(\bs{\e})} & = \frac{p_1(\ell_1)}{p_1(\e_{n})}  \left[\frac{p(\up | \up)}{p(\down | \down)}\right]^{n_{\up\up}(\bs{\ell}) - n_{\down\down}(\bs{\ell})} \label{eq:ratio}
\end{align}
where $n_{\ell\ell'}(\bs{\ell})$ is the number of times trans-transition $\ell \to \ell'$ occurs along the path. 

\begin{figure}[t]
\begin{center}
\includegraphics[width=\columnwidth]{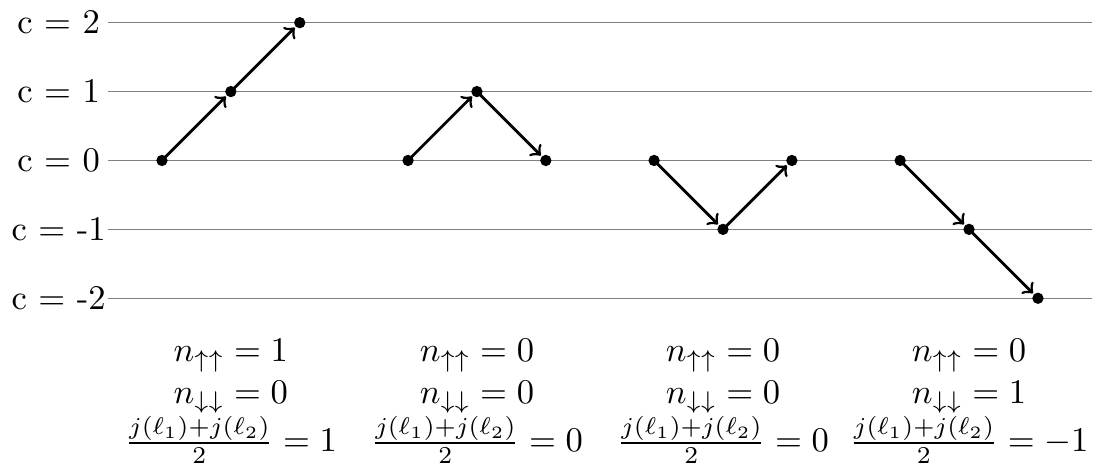}
\caption{
    The diagram shows how boundary terms contribute to the current by displaying trajectories of length \( n = 2\), uniquely formed by boundary terms. Longer trajectories will only change the values of \( n_{\uparrow\uparrow}\) and \(n_{\downarrow\downarrow}\) and Eq.\,(\ref{eq:c2}) will hold.
}
\label{fig:current}
\end{center}
\end{figure}

\paragraph{Currents and the fluctuation relation.} Letting $j (\ell) := \delta_{\ell,\uparrow} - \delta_{\ell,\downarrow}$ be the instantaneous current, signaling when a transition occurs, we focus on the cumulated current (or charge)
\begin{align}
\cur(\bs{\ell}) := \sum_{k = 1}^n j (\ell_k) = n_{\up}(\bs{\ell}) -  n_{\down}(\bs{\ell}). \label{eq:c1}
\end{align}
where $n_\ell$ is the number of times transition $\ell$ has been performed along the process. Notice that, it can only take values $\{-n, -n + 2, \ldots , n- 2, n\}$, and that it is anti-symmetric by time reversal, $c(\overline{\bs{\ell}}) = - c(\bs{\ell})$. Importantly, we can also express it in terms of the trans-transition numbers
$n_{\ell\ell'}$ as
\begin{align}
\cur(\bs{\ell}) = n_{\up\up}(\bs{\ell}) - n_{\down\down}(\bs{\ell}) + \frac{j(\ell_1) +  j(\ell_n)}{2}.  \label{eq:c2}
\end{align}
The first term is due to the fact that occurrences of $\up\down$ and $\down\up$ always reset the current to its initial value, and therefore only self-loops contribute to it. The second boundary term is less intuitive, and is explained in Fig.\,\ref{fig:current}.

The central result of this Letter is found${}^6$ by plugging this latter expression into  Eq.\,(\ref{eq:ratio}) and summing over all intermediate transitions $\ell_2,\ldots,\ell_{n-1}$. After standard manipulations we find the fluctuation relation for the joint  probability of the charge and of the first and last transitions
\begin{align}
p_n(\cur, \ell_1,\ell_n) & = p_n(-\cur, \e_n, \e_1 ) \exp \left[ f^\varnothing  \cur + u(\e_n) - u(\ell_1) \right] \label{eq:fluctuation relationjoint}
\end{align}
where, given an arbitrary constant \(v\), the effective force $f^\varnothing$ \cite{polettini2019effective} and the effective potential \(u\) are given by:
\begin{align}
f^\varnothing := \log \frac{p(\up | \up)}{p(\down|\down)} \label{eq:effaff}, \qquad u(\ell) : = \frac{j(\ell) f^\varnothing}{2} -  \log p_1(\ell) + v.
\end{align}
Parametrizing the visible rates by the principle of local detailed balance $r(2|1) / r(1|2) = \exp \beta \varepsilon$ with $\beta$ a tunable inverse temperature (in units of Boltzmann's constant) and of a fixed energy increment $\varepsilon$, the effective affinity can be shown to be given by $f^\varnothing = (\beta - \beta^\varnothing) \varepsilon$ where $\beta^\varnothing$ is the stalling value that makes the visible current vanish on average \cite{polettini2022phenomenological,neri2022extreme}. Thus, operationally, provided $\varepsilon$ is known from microphysical considerations, all one has to do to obtain $f^\varnothing$ is to tune the temperature to the stalling state and measure the difference.
\begin{figure}[t]
\includegraphics[width=.9\columnwidth]{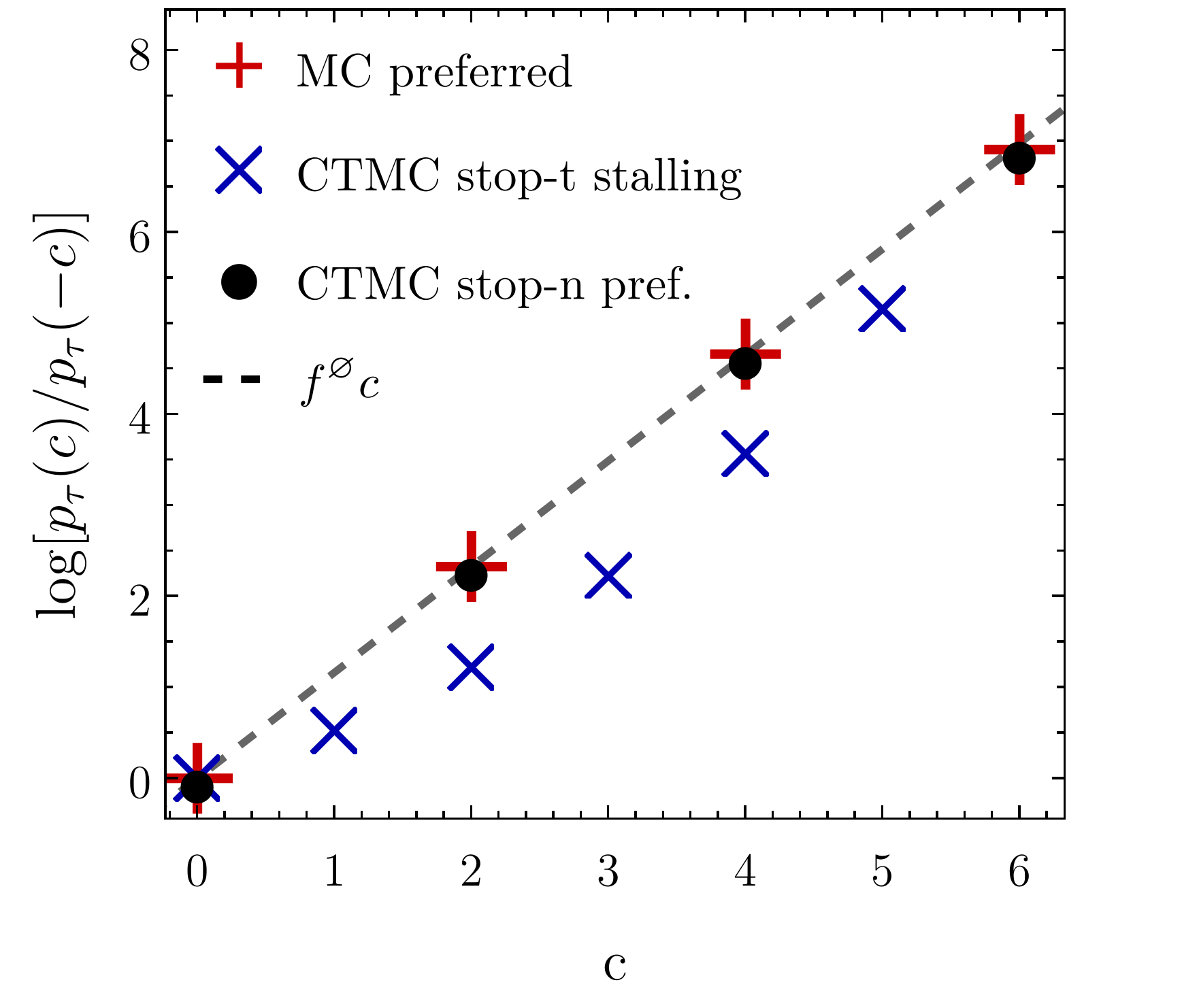}
\caption{Log-ratio $\log p_\tau(+c) / p_\tau(-c)$, $c \geq 0$, for different stopping times and processes in a four-state system: state-space continuous-time Markov chain at stopping-$t$ sampled from the stalling distribution (CTMC stop-t stalling) and at stopping-$n$ sampled from the preferred distribution (CTMC stop-n pref.), and the transition-space Markov chain also sampled from the preferred distribution (MC preferred). Only the latter two satisfy the fluctuation relation. More details of the simulation in the Supplemental Material${}^7$.
}
\label{fig:scgf}
\end{figure}

The potential $u$ can be made to vanish by selecting as preferred initial distribution (marked $^\ast$)
\begin{align}\label{eq:preferred}
p_1^\ast(\ell) \propto  p(\ell|\ell)
\end{align}
with $v$ chosen to fix the normalization. In view of Eq.\,(\ref{eq:init}), a state-space precursor of this distribution is $q^\ast_0(1)  \propto p(\down\vert\down)$, $q^\ast_0(2) \propto p(\up\vert\up)$, else zero. When the potential vanishes one can further marginalize for the current in Eq.\,(\ref{eq:fluctuation relationjoint}) by summing over $\ell_1$ and $\ell_n$ to obtain our central result, the fluctuation relation at finite number of total visible transitions
\begin{align}\label{eq:FT}
    \frac{p_n(c)}{p_n(-c)} = \exp f^\varnothing c.
\end{align}

Figure~\ref{fig:scgf} illustrates the validity of the fluctuation relation for the state-space continuous-time Markov chain at stopping-\(n\), prepared from $\vec{q}_0^{\,\ast}$, and the transition-space Markov chain, prepared from $\vec{p}_1^{\,\ast}$. It also displays the failure of the fluctuation relation for the continuous-time Markov chain at clock time, prepared from the so-called stalling distribution, which is the best candidate for the preferred initial distribution given that it satisfies the integral fluctuation relation $\langle \exp f^\varnothing c \rangle = 1$ \cite{polettini2019effective}. The fluctuation relation also holds asymptotically, beyond the finite values of charge explored in these numerical simulations, as revealed by a symmetry in the stationary scaled cumulant generating function${}^8$ \cite{andrieux2007fluctuation,polettini2014transient}.

\paragraph{Discussion and conclusions.}

Three different stochastic processes are at play in this work. We started with a continuous-time Markov chain sampled from some distribution $\vec{q}_0$ and evolving by generator $R$ up until final clock time $t$. The second is in all identical, but, as already suggested in Refs.\, \cite{kiukas2015equivalence,budini2014fluctuating}, and explored for run-and-tumble particles with fixed number of jumps \cite{Mori_2021, Gradenigo_2019, PhysRevE.103.062134}, it runs until $n$ transitions are observed. Finally, we introduced a discrete-time Markov chain in transition space with (trans-)transition matrix $P$ sampled from $\vec{p}_1$. Other internal notions of time have been considered in Refs.\,\cite{pigolotti2017generic, pietzonka2022classical}

The latter two produce the same current statistics, as evidenced by red crosses and black dots in Fig.\,\ref{fig:scgf}, thus confirming our second main point, viz. the appropriateness of the concept of Markov chains in the space of transitions for the study of transitions at their own beat. An additional statistical advantage of working at fixed number of transitions is that no events of charge $c > n$ can possibly occur, thus making the distribution of compact support -- while at clock time events of rare activity produce noise in the tails that are hard to characterize in simulations.

The results presented here hold only if every visible transition, $\up$ or $\down$ arises from a unique and well-defined transition in state space. Otherwise, if two transitions produce the same signal, the so-called renewal property is lost, and correspondingly the time-series of transitions is not Markovian (i.e. there will be a difference in probability between $p(\up | \down , \down)$ and $p(\up | \down, \up)$ etc.). A different question is whether the fluctuation relation holds for several independent transitions. Work in progress by the Authors suggests that monitoring the number of occurrences of the visible transitions is not enough, and that one needs to additionally consider cross-information.

Many facts that are true for processes in state space may not be true in transition space, and vice-versa. One example is the notion of time-reversal, where the requirement of also flipping the transition's direction resembles the time-reversal of the momentum variable in underdamped Langevin dynamics. Thus, as anticipated in Ref.\,\cite{van2022thermodynamic}, this formalism may serve as a candidate for ``second-order'' Markov processes as the basis for transition-based coarse-graining \cite{harunari2022learn,van2022thermodynamic, martinez2019inferring,hartich2021emergent,hartich2021comment,bisker2022comment}.

Our central result, Eq.~\eqref{eq:FT}, is obtained from the combination of the formalism in Refs.~\cite{harunari2022learn, van2022thermodynamic} and the introduced notion of stopping-\(n\), further assessing the fluctuating nature of single currents. The present fluctuation relation holds even when the fundamental set is not fully accessible, composing a relevant observation for understanding the thermodynamic consistency of currents flowing in the vast class of partially observed systems. Another connection between these works is the entropic interpretation: Usually, exponents of fluctuation relations are measures of dissipation. Indeed, the effective affinity times the visible current \( f^\varnothing c\) bounds the entropy production of a stochastic process from below, as shown in \cite{harunari2022learn, van2022thermodynamic}.

From the present fluctuation relation and the definition of Markov processes in transition space, thermodynamic considerations beyond dissipation inference might arise; for example, an associate thermodynamic uncertainty relation \cite{barato2015thermodynamic, gingrich2016dissipation, PhysRevLett.123.110602}, the usage of Martingale theory \cite{roldan2022martingales} to explore the beat of a current as a random stopping time, connections to fluctuating heat and work that are now accessible at small-scales \cite{blickle2012realization, PhysRevLett.109.180601}, the measurement of effective affinities from current statistics \cite{note:affinity}, and the usage of discrete-time processes to efficiently assess some details of continuous-time Markov chains${}^7$.

\paragraph*{Acknowledgments.}

Emanuele Penocchio for helping out with the introduction, Gianmaria Falasco for suggesting the connection of the survival matrix to large deviations, Danilo Forastiere for sharing ideas, Pedro Portugal for helping with computational resources, Michela Bernini for the illustration in Fig.~\ref{fig:fig1}. PH thanks Massimiliano Esposito for hospitality in his group. The research was supported by the National Research Fund Luxembourg (project CORE ThermoComp C17/MS/11696700), by the European Research Council, project NanoThermo (ERC-2015-CoG Agreement No. 681456), and by grants \#2017/24567-0 and \#2020/03708-8, S\~ao Paulo Research Foundation (FAPESP).

\nocite{note:SMrefs, andrieux2004fluctuation, touchette2009large}

\bibliographystyle{ieeetr}
\bibliography{biblio}

\begin{thebibliography}{10}
\bibitem{note:affinity}
The affinity can be measured as the slope of \(\log p(c) / p(-c)\) when \(c\)
  satisfies a fluctuation relation, the same holds for the effective affinity
  between two hidden states.
\end{thebibliography}


\widetext
\begin{center}
\textbf{\large Supplemental Material: Beat of a current}
\end{center}
\setcounter{equation}{0}
\setcounter{figure}{0}
\renewcommand{\theequation}{S\arabic{equation}}
\renewcommand{\theHequation}{S\arabic{equation}}
\renewcommand{\thefigure}{S\arabic{figure}}
\renewcommand{\theHfigure}{S\arabic{figure}}

\sml{genfun}{Derivation of survival probability}

We propose a simple derivation based on large deviations. It is well known \cite{andrieux2007fluctuation} that the moment generating function of the probability $p_t(\bs{n}|y)$ of observing $\bs{n} = (n_{\up}, n_{\down})$ transitions $\up, \down$ at time $t$, starting from $x$, is given by
\begin{align}
\sum_{\bs{n} \in \mathbb{Z}^2} e^{\bs{\lambda} \cdot \bs{n}} p_t(\bs{n}|x) = \sum_y \left[\exp t R^{\bs{\lambda}}\right]_{y,x} 
\end{align}
where the tilted matrix $R^{\bs{\lambda}}$ is obtained by biasing $R$ in the off-diagonal entries $r^{\bs{\lambda}}(1|2) := r(1|2) \exp \lambda_{\up}$ and $r^{\bs{\lambda}}(2|1) := r(2|1) \exp \lambda_{\down}$. In the limit $\bs{\lambda} \to -\infty$ we obtain that the only term surviving in the left-hand side is the survival probability that, starting from state $x$, the system has not performed any visible transition up to time $t$ \cite{sekimoto2021derivation,harunari2022learn}:
\begin{align}
p_t(\bs{0}|x) = \sum_y \left[\exp t \R \right]_{y,x}.
\end{align}

\sml{negative}{Negativity of eigenvalue of $\R$ with largest real-part}

Consider $M^{\bs{\lambda}} = R^{\bs{\lambda}} + \max_y r(y) I$, where $R^{\bs{\lambda}}$ is the tilted matrix as defined  in the previous paragraph, and $I$ is the identity. Then $M^{\bs{\lambda}}$ is a non-negative irreducible matrix, and $M^{\bs{0}}$ has spectral Perron root $\max_y r(y)$.

It is well-known that the Perron root of a non-negative irreducible matrix is a non-decreasing function of all its components \cite{suzumura1983perron}. Let us now prove that it actually depends on $\bs{\lambda}$, so that it has to be a strictly decreasing function of $-\bs{\lambda}$. Intuitively, given that the Perron root of $M^{\bs{\lambda}}$ is $g(\bs{\lambda}) + \max_y r(y)$ with $g(\bs{\lambda})$ the scaled-cumulant generating function of the stationary fluxes along $1\leftrightarrow 2$, and because we assumed that $1\leftrightarrow 2$ belongs to some cycle, its fluxes' cumulants cannot all be vanishing, and therefore $g(\bs{\lambda})$ does depend on $\bs{\lambda}$ (this would not be the case if $1\leftrightarrow 2$ was supported on a leaf). A more formal argument could be made based on the fact that the coefficients of the characteristic polynomial of $M^{\bs{\lambda}}$ depend on all matrix elements belonging to some cycle \cite[Eq.\,50]{andrieux2004fluctuation}, \cite[Sec. 5.7]{polettini2019effective}.

But since $\R + \max_y r(y) I$ is obtained from $M^{\bs{0}} = R + \max_y r(y) I$ by lowering the positive entries $R_{1,2}$ and $R_{2,1}$ to zero, its Perron root must be smaller than $\max_y r(y)$, and therefore all eigenvalues of $\R$ have negative real part.

\newpage

\sml{transtrans}{Trans-transition probabilities}

Using the same reasoning as for main text's Eq.\,(6), we explicitly obtain:
\begin{align}
\label{eq:transtrans}
\begin{split}
p(\up|\up) & = - r(1|2) \, [\R^{-1}]_{2,1} \\
p(\up|\down) & = - r(1|2) \, [\R^{-1}]_{2,2} \\
p(\down|\up) & = - r(2|1) \, [\R^{-1}]_{1,1}  \\
p(\down|\down) & = - r(2|1) \, [\R^{-1}]_{1,2} 
\end{split}
\end{align}

\sml{chains}{Expansion in terms of induced Markov chains}

Here, we prove Eq.~(7) from the main text, which is used to build intuition on the role of \(S^{-1}\) in trans-transition probabilities, reproduced here for convenience:
\begin{align}
- [\R^{-1}]_{x,x_0} = \frac{1}{r(x)} \sum_{\bs{x} \;:\; x_0 \rightsquigarrow x} \theta(\bs{x}) \, p(\bs{x} | x_0), \label{eq:pathhidden}
\end{align}
where the sum runs over all trajectories, of any length, that go from $x_0$ to $x$.

First, we introduce the transition matrix $Q$ of the induced Markov chain $\bs{x}$, with entries $Q_{x,y} = r(x|y) / r(y)$ and $Q_{x,x} = 0$. Indeed, the off-diagonal element $Q_{x,y} = p( x_{k+1} = x | x_k = y)$ is the probability that the next visited state is $x$, given $y$. Also, we define the survival transition matrix as $M := Q \circ T$ with $M_{1,2} = 0 = M_{2,1}$.

Rearranging $\R = (BA^{-1} - I)A$ where $-A$ and $B$ are respectively its diagonal and off-diagonal parts, and $I$ is the identity. Notice that $BA^{-1} = M$, whose spectral radius can be easily shown to be $\rho(M) < 1$ using the same arguments of Section~2.
Then the Neumann operator geometric series $\sum_{m} M^m$ converges to $(I - M)^{-1}$, yielding

\begin{align}
- [\R^{-1}]_{x,x_0} = \left[A^{-1} \sum_{m \in \mathbb{N}} M^m \right]_{x,x_0} = \frac{1}{r(x)} \sum_{m \in \mathbb{N}} [M^m]_{x,x_0}
\end{align}

We now expand the matrix powers above as
\begin{align}
\left[M^m\right]_{x,x_0} = \sum_{x_1,\ldots,x_m} \delta_{x_m,x} \prod_{k = 0}^{m-1} Q_{x_{k+1},x_k} T_{x_{k+1},x_k}
\end{align}
The conclusion follows after recognizing
\begin{align}
\theta(\bs{x}) & = \prod_{k = 0}^{m-1} T_{x_{k+1},x_k} \\
p(\bs{x} | x_0) & = \prod_{k = 0}^{m-1} Q_{x_{k+1},x_k}.
\end{align}

\sml{normalization}{Normalization of initial distribution}

We want to confirm that $p_1(\ell)$ is normalized. We have
\begin{align}
-\sum_{\ell} r(\ell)  [ \R^{-1} ]_{\mathtt{s}(\ell),x} & =  \sum_{y,z} (\R-R)_{y,z}  [ \R^{-1} ]_{z,x} \nonumber \\
 & =  \sum_{y,z} \R_{y,z}  [ \R^{-1} ]_{z,x} \nonumber \\
 & =  \sum_{y} I_{y,x} \nonumber \\
 & = 1
\end{align}
where we used the fact that columns of $R$ add up to zero. Therefore from main text's Eq.\,(10) 
\begin{align}
\sum_{\ell} p_1(\ell) = \sum_{x} q_0(x) = 1.
\end{align}

\sml{fluctuation}{Derivation of the fluctuation relation}

From the probability of a path in transition-space $p(\bs{\ell})$ we can introduce the joint probability of a path and the value of its charge
\begin{equation}
    p(\cur, \bs{\ell}) = \delta_{\cur, \cur(\bs{\ell})}p(\bs{\ell}),
\end{equation}
which is normalized since $\sum_{\cur, \bs{\ell}} p(\cur, \bs{\ell}) = \sum_{\bs{\ell}} p(\bs{\ell}) = 1$.

The probability of a path can be expressed in terms of the probability of the number of its pairs of transitions. Marginalizing over the bulk transitions we obtain

\begin{align}
    p(\cur,\ell_1, \ell_n) &= \sum_{\ell_2,\ldots,\ell_{n-1}} p(\cur, \bs{\ell}) = \sum_{\ell_2,\ldots,\ell_{n-1}} \delta_{\cur, \cur(\bs{\ell})} p(\bs{\ell}) \nonumber\\
    &=\sum_{\ell_2,\ldots,\ell_{n-1}} \delta_{\cur, \cur(\bs{\ell})} p_1(\ell_1) \prod_{m=2}^n p(\ell_m \vert, \ell_{m-1}) \nonumber\\
    &= p_1(\ell_1) \sum_{\ell_2,\ldots,\ell_{n-1}} \delta_{\cur, \cur(\bs{\ell})} [ p(\up \vert \up)]^{n_{\up\up}(\bs{\ell})} [ p(\down \vert \down)]^{n_{\down\down}(\bs{\ell})}  [ p(\up \vert \down)]^{n_{\up\down}(\bs{\ell})} [ p(\down \vert \up)]^{n_{\down\up}(\bs{\ell})}
\end{align}
whereas for the time-reversed the effective force $f^\varnothing$ can be cast to express pairs of transitions of $\bar{\bs{\ell}}$ in terms of $\bs{\ell}$'s pairs:
\begin{align}
    p(-\cur,\bar{\ell}_1, \bar{\ell}_{n}) &= \sum_{\bar{\ell}_2,\ldots,\bar{\ell}_{n-1}} p(-\cur,\bar{\bs{\ell}}) = \sum_{\bar{\ell}_2,\ldots,\bar{\ell}_{n-1}} \delta_{-\cur, \cur(\bar{\bs{\ell}})} p(\bar{\bs{\ell}})\nonumber\\
    &=\sum_{\ell_2,\ldots,\ell_{n-1}} \delta_{\cur, \cur(\bs{\ell})} p_1(\bar{\ell}_n) \prod_{m=2}^n p(\bar{\ell}_{m-1} \vert, \bar{\ell}_{m}) \nonumber\\
    &= p_1(\bar{\ell}_n) \sum_{\ell_2,\ldots,\ell_{n-1}} \delta_{\cur, \cur(\bs{\ell})} [ p(\up \vert \up)]^{n_{\up\up}(\bar{\bs{\ell}})} [ p(\down \vert \down)]^{n_{\down\down}(\bar{\bs{\ell}})} 
    [ p(\up \vert \down)]^{n_{\up\down}(\bar{\bs{\ell}})} [ p(\down \vert \up)]^{n_{\down\up}(\bar{\bs{\ell}})} \nonumber\\
    &= p_1(\bar{\ell}_n)\sum_{\ell_2,\ldots,\ell_{n-1}} \delta_{\cur, \cur(\bs{\ell})} [ p(\down \vert \down)e^{f^\varnothing}]^{n_{\down\down}(\bs{\ell})} [ p(\up \vert \up) e^{-f^\varnothing}]^{n_{\up\up}(\bs{\ell})}  [ p(\up \vert \down)]^{n_{\up\down}(\bs{\ell})} [ p(\down \vert \up)]^{n_{\down\up}(\bs{\ell})} \nonumber\\
    &= p_1(\bar{\ell}_n)  e^{-f^\varnothing( \cur - [j(\ell_1)+j(\ell_n)]/2 )} \frac{p(\cur, \ell_1, \ell_n)}{p_1(\ell_1)}.
\end{align}
Here, we describe the steps taken in the above derivation. In the second line we used that the sum over all possible paths equals the sum for the time-reversed paths $\sum_{\ell_m} = \sum_{\bar{\ell}_m}$ and that reversing the path changes the sign of the charge $\delta_{c,c(\mathbf{\ell})} = \delta_{-c,c(\mathbf{\bar{\ell}})}$; in the fourth line we used that the number of pairs of repeated transitions swap under time reversal, $n_{\uparrow\uparrow}(\bar{\ell}) = n_{\downarrow\downarrow}(\ell)$ and $n_{\downarrow\downarrow}(\bar{\ell}) = n_{\uparrow\uparrow}(\ell)$, while alternated do not, $n_{\uparrow\downarrow}(\bar{\ell}) = n_{\uparrow\downarrow}(\ell)$ and $n_{\downarrow\uparrow}(\bar{\ell}) = n_{\downarrow\uparrow}(\ell)$; in the fifth line we used the value of effective affinity to get $e^{f^\varnothing} = p(\uparrow\vert\uparrow) /p(\downarrow\vert\downarrow)$ and main text's Eq. (16) $c(\ell) = n_{\uparrow\uparrow}(\ell) - n_{\downarrow\downarrow}(\ell) + [j(\ell_1)+j(\ell_n)]/2$.

Notice that $j(\ell_n) = -j(\bar{\ell}_n)$ and, introducing the potential $u(\ell):= f^\varnothing j(\ell)/2 - \log p_1(\ell)$, above expression simplifies to the joint fluctuation relation (FR) of charge and boundary transitions
\begin{align}
    \frac{p(\cur,\ell_1, \ell_n)}{p(-\cur,\bar{\ell}_1, \bar{\ell}_n)} =  \exp\lbrace f^\varnothing \cur +u(\bar{\ell}_n) - u(\ell_1) \rbrace.
\end{align}

\sml{simulations}{Details of simulations}

We validate our results by performing numerical experiments where we generate a continuous-time Stochastic Process on a minimal 4 states network analogous to the one sketched in second page of the main text with transition rates $r(2|1) = 5$, $r(1|2) = 4$, $r(4|1) = 2$, $r(1|4) = r(2|3) = r(3|1) = r(3|2) = r(3|4) = r(4|3) = 1$ together with $r(1|3) = 0$ to also include a hidden irreversible transition, illustrating that the framework also accommodates irreversible transitions. With this choice of transition rates, the effective affinity is \(f^\varnothing \approx 1.163\). Three different experiments are performed with this choice of rates:

\begin{enumerate}
    \item [i] (CTMC stopping-\(n\)) The simulation makes use of the Doob-Gillespie algorithm, i.e. once the initial state \(y\) is chosen according to a distribution \(q_0\) at time \(t = 0\), the system jumps after a time drawn from an exponential distribution with parameter \(r(y) = \sum_{x \neq y} r(x \vert y)\) to state \(x\) with probability \(r(x \vert y)/ r(y)\). This procedure is repeated and the charge \(c\) is monitored with the convention that \(1 \gets 2\ (2 \gets 1)\) increases (decreases) it by 1. After \(n = 6\) visible transitions the algorithm stops, the total current is registered, and all the procedure is repeated with the same initial probability for \(10^8\) times. We collect the outcomes of \(c\) in a histogram that corresponds to an estimation of the probability \(p(c)\).
    
    \item [ii] (CTMC stopping-\(t\)) The same procedure above is used for the fixed-\(t\) process where the total current \(c\) of a single realization of the experiment is now evaluated up to time \(t\). For a fair comparison between stopping-\(n\) and stopping-\(t\), we first record the duration \(s\) of every stopping-\(n\) trajectory and then use the average \(\langle s \rangle\) as stopping-\(t\) criterion.

    \item [iii] (MC stopping-\(n\)) The last numerical experiment consists of a simulation of a discrete-time Markov chain evolving according to the trans-transition matrix [Eq. (8) in the main text], stopping after \(n = 6\) steps are performed, representing the occurrence of 6 visible transitions. Each occurrence of transitions \( \uparrow \ (\downarrow)\) increases (decreases) the total charge \(c\) by 1.
\end{enumerate}

Experiments (i) and (iii) generate exactly the same statistics for the total current \(c\). In Figure 3 of the main text, (iii) starts from the preferred distribution \(p_1^*(\ell)\) and (i) starts from a state space configuration \(q_0^*(x)\) compatible with the preferred distribution, satisfying the FR at all values of \(c\). Conversely, experiment (ii) is used to prove the fact that a fluctuation relation
does not hold at stopping-\(t\) when observing a single current on a network with more than one cycle, since even from the best candidate initial state, namely the stalling state, no linearity is present.

As an additional remark, updating the transition space Markov chain as in (iii) marginalizes away the hidden transitions and states, as
well as the state sojourn times, thus representing a computational advantage for assessing integrated visible currents’ statistics ``at the beat of a current''. For the referred system, transition space Markov chain’s computational time is smaller by one order of magnitude when compared to CTMC simulations.

\sml{asymp}{Asymptotic behavior}

The main results were illustrated in numerical simulations that account for finite values of charge \(c\). As some FRs only hold asymptotically, we raise the hypothesis that the FR for a single current might hold for the stationary charge \(c\) at large times and/or large number of visible transitions' occurrences. As already used in the derivation of FRs for currents \cite{andrieux2007fluctuation}, a symmetry in the cumulant generating function reveals the FR at large times.

Let the stationary scaled cumulant generating function (SSCGF) be
\begin{equation}\label{eq:g_def}
    g_{\tau} (\lambda) = \lim_{\tau\to\infty} \frac{1}{\tau} \log \langle e^{\lambda c} \rangle_\tau,
\end{equation}
where \(\tau\) can be either the stopping criteria \(t\) or \(n\), and the charge \(c\) depends on \(\tau\). The parameter \(\lambda\) is the counting field, and derivatives in terms of \(\lambda\) provide the cumulants of the current. For the case of stopping-\(t\), it can be obtained as the largest eigenvalue of the tilted rate matrix \cite{touchette2009large}:
\begin{equation}
    \tilde{R}_{x,y}(\lambda) := \begin{cases} R_{1,2} e^{\lambda} & \text{if } x=1 \text{ and } y = 2 \\ R_{2,1} e^{-\lambda} & \text{if } x=2 \text{ and } y = 1 \\ R_{x,y} & \text{otherwise} \end{cases} 
\end{equation}
Now, for the case of stopping-\(n\), it is given by the logarithm of the largest eigenvalue of the tilted trans-transition matrix:
\begin{equation}
    \tilde{P} := \begin{pmatrix} p(\up|\up) e^\lambda & p(\up|\down) e^\lambda \\ p(\down|\up) e^{-\lambda} & p(\down|\down) e^{-\lambda} \end{pmatrix}
\end{equation}
since the process is discrete in time.

When the FR is satisfied, the SSCGF presents the following symmetry:
\begin{equation}\label{eq:fr}
    g_{\tau} (\lambda) = g_{\tau} (-f - \lambda),
\end{equation}
which has a distinct sign compared to other references (e.g. \cite{andrieux2007fluctuation, polettini2014transient}) due to differences in the definition of SSCGF.

The cumulants generated by \(g_{n}\) have to be scaled by \(r^\infty\) to be compared to those obtained by the usual \(g_t\), since \(r^\infty\) is the rate of visible transitions, viz., the ``average beat''. The statistics will in general be different, which is clear since one satisfies an FR and the other does not; however, it can be shown that the first cumulants match: \(g_{t}'(0) = r^\infty g_n' (0)\).  It is evident from the different dimensions of \(\tau\) in Eq.~\eqref{eq:g_def}, where \(\tau = t\) provides cumulants per unit of time and \(\tau = n\) per unit of transitions.

\begin{figure}
    \centering
    \includegraphics[width=.48\textwidth]{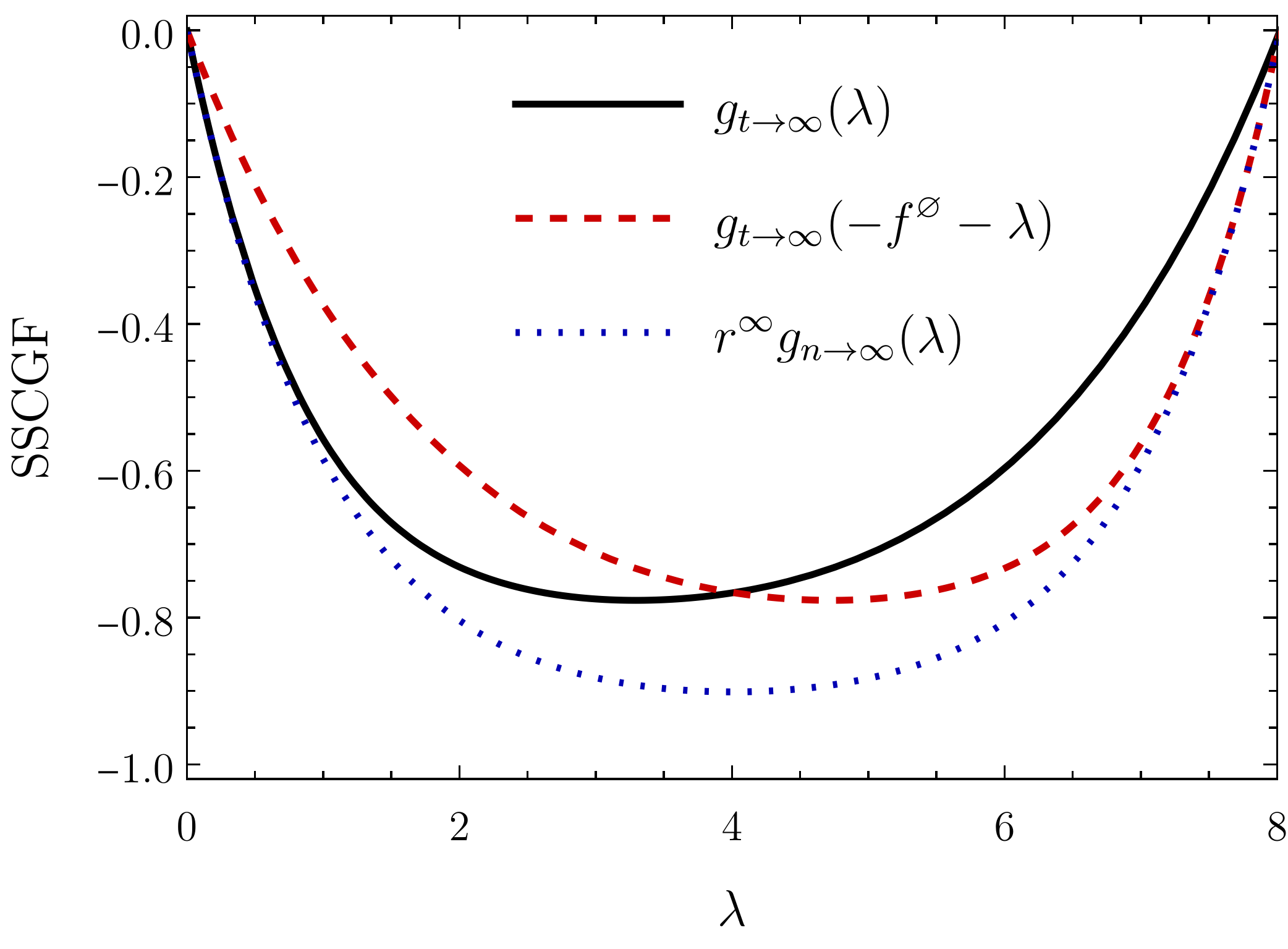}
    \caption{Charge's scaled cumulant generating function at long times \(g_{t}\), and at at large number of transitions \(g_{n}\). Only the second displays the fluctuation symmetry $\lambda \to - f^\varnothing - \lambda$. The system is a four-state model with two internal currents and transition rates \(R_{1,2} = R_{2,3} = R_{3,4} = R_{4,1} = R_{4,2} = 1\), \( R_{1,4} = R_{2,1} = R_{3,2}^{-1} = R_{4,3}^{-1} = e^4\). The visible transitions are \(\up = 1 \gets 2\) and \(\down = 1\to 2\). In this case, \(f^\varnothing = -8.01848 \).}
    \label{fig:sscgf}
\end{figure}

Figure \ref{fig:sscgf} shows that, for a given system, the SSCGF is non-symmetric for stopping-\(t\) even at long times, illustrating the absence of a FR. Note that other values of \(f^\varnothing\) still won't satisfy the symmetry. However, for the case of stopping-\(n\), the FR is revealed by the symmetric function \(g_{n} (\lambda) = g_{n}(-f^\varnothing - \lambda) \). In accordance to the \(c \gg 1\) behavior of the FR derived in the main text
\begin{equation}
    \frac{ p_n(\cur, \ell_1,\ell_n)}{p_n(-\cur, \e_n, \e_1 )} = \exp \left[ f^\varnothing  \cur + u(\e_n) - u(\ell_1) \right] \asymp \exp \left( f^\varnothing  \cur \right)
\end{equation}

Finally, we observe that the SSCGF vanishes when \(f^\varnothing + \lambda = 0\). From the definition \eqref{eq:g_def} and the fact that the logarithm is bijective, it is a manifestation of the integral fluctuation theorem (IFT)
\begin{equation}
    \langle e^{-f^\varnothing c} \rangle = 1,
\end{equation}
and the IFT itself is a consequence of the FR in \eqref{eq:fr}.

\end{document}